\documentclass[final,5p,times,twocolumn]{elsarticle}

\usepackage{graphics}
 \usepackage{graphicx}
 \usepackage{epsfig}
\usepackage{amssymb}
\usepackage{lineno}

 \usepackage{epstopdf}
\usepackage{amsmath}
\usepackage{subfigure}
\usepackage{tikz}
\usepackage{lipsum}

\newcommand\copyrighttext{%
  \footnotesize Published in Nuclear Instruments and Methods in Physics Research A 740 (2014) 165-172, DOI:10.1016/j.nima.2013.10.092}
\newcommand\copyrightnotice{%
\begin{tikzpicture}[remember picture,overlay]
\node[anchor=south,yshift=10pt] at (current page.south) {\fbox{\parbox{\dimexpr\textwidth-\fboxsep-\fboxrule\relax}{\copyrighttext}}};
\end{tikzpicture}%
}

\journal{Nuclear Instruments and Methods in Physics Research A }

\begin{document}

\begin{frontmatter}



\title{A plasma wakefield acceleration experiment using CLARA beam}


\author[a,b]{G. Xia}
\author[c]{D. Angal-Kalinin}
\author[c]{J. Clarke}
\author[d]{J. Smith}
\author[e]{E. Cormier-Michel}
\author[c]{J. Jones}
\author[c]{P. H. Williams}
\author[c]{J. W. Mckenzie}
\author[c]{\\B. L. Militsyn}
\author[a,b]{K. Hanahoe}
\author[a,b]{O. Mete}
\author[b,f]{A. Aimidula}
\author[b,f]{C. Welsch}

\address[a] {School of Physics and Astronomy, University of Manchester, Manchester, United Kingdom}
\address[b]{The Cockcroft Institute, Sci-Tech Daresbury, Daresbury, Warrington, United Kingdom}
\address[c]{STFC/ASTeC, Daresbury, Warrington, United Kingdom}
\address[d]{Tech-X UK Corporation, Daresbury Innovation Centre, Warrington, United Kingdom}
\address[e]{Tech-X Corporation, Boulder, Colorado, USA}
\address[f]{The University of Liverpool, Liverpool, United Kingdom}

\begin{abstract}
We propose a Plasma Accelerator Research Station (PARS) based at proposed FEL test facility CLARA (Compact Linear Accelerator for Research and Applications) at Daresbury Laboratory. The idea is to use the relativistic electron beam from CLARA, to investigate some key issues in electron beam transport and in electron beam driven plasma wakefield acceleration, e.g. high gradient plasma wakefield excitation driven by a relativistic electron bunch, two bunch experiment for CLARA beam energy doubling, high transformer ratio, long bunch self-modulation and some other advanced beam dynamics issues. This paper presents the feasibility studies of electron beam transport to meet the requirements for beam driven wakefield acceleration and presents the plasma wakefield simulation results based on CLARA beam parameters. Other possible experiments which can be conducted at the PARS beam line are also discussed.

\end{abstract}
\end{frontmatter}
\copyrightnotice

\section{Introduction}
\label{}
Plasma accelerators utilize the breakdown medium ``plasmas'' as the accelerating structures and therefore avoid the further material breakdown limit posed by the conventional accelerating structures, e.g. copper or niobium RF cavities. In principle, the wavebreaking field can reach 100$\,$GV/m for a plasma with density of $10^{18}\,cm^{-3}$. Plasmas are therefore the ideal medium to sustain a very large electric field for particle beam acceleration.

Plasma based accelerators have achieved tremendous progress in recent decades \cite{esarey,joshi}. With the advances in laser technology, especially with the introduction of the Chirped Pulse Amplification (CPA) \cite{strickland}, the cutting edge laser facility nowadays can routinely reach a few hundred Terawatts ($10^{12}\,$Watts) or even Petawatts ($10^{15}\,$Watts) peak power with very short pulse length (e.g., tens of femtoseconds). Employing such a laser pulse as drive beam, laser wakefield accelerators (LWFA) can achieve $\sim$GeV level electron beam within a few centimeter plasma channel and with an energy spread of only a few percent \cite{leemans1, wang1}. This scheme therefore has enormous potential for use in future compact light source \cite{cipiccia} or particle colliders \cite{leemans2}. For the electron beam driven plasma wakefield accelerator (PWFA), the experiments conducted by a group of scientists from UCLA/USC/SLAC collaboration at the FFTB at SLAC, have successfully demonstrated energy doubling for the SLC $42\,$GeV electron beam \cite{blumenfeld}. The resulting accelerating gradient in this experiment is about three orders of magnitude higher than the accelerating field of the SLC Linac \cite{muggli1}. Compared to the intrinsic limitations posed by LWFA scheme (e.g. depletion, diffraction and dephasing), the relativistic charged particle beam can in principle propagate in a plasma for a long distance due to large beam beta function (beta function of the beam is equivalent to the Rayleigh length of a laser beam). Therefore the beam driven PWFA has already attracted worldwide interests. It can potentially accelerate a particle beam to high energy due to its high accelerating gradient and relatively long acceleration length. In this paper we propose a new research facility PARS (Plasma Accelerator Research Station) for studying plasma wakefield acceleration at the CLARA facility based at Daresbury Laboratory.

This paper is structured as follows: in section 2, the CLARA facility and the design strategy for the PARS beam line are introduced. Section 3 elaborates on the electron beam tracking results through the CLARA/PARS beam line. The beam parameter settings at the CLARA/PARS are given in section 4. The detailed particle-in-cell (PIC) simulation results on plasma wakefields driven by the electron beams at various plasma densities are presented in section 5. Section 6 discusses some other research topics (issues), which can be investigated at the PARS beam line as well.

\begin{figure}[ht] 
\centering
\includegraphics[width=0.45\textwidth]{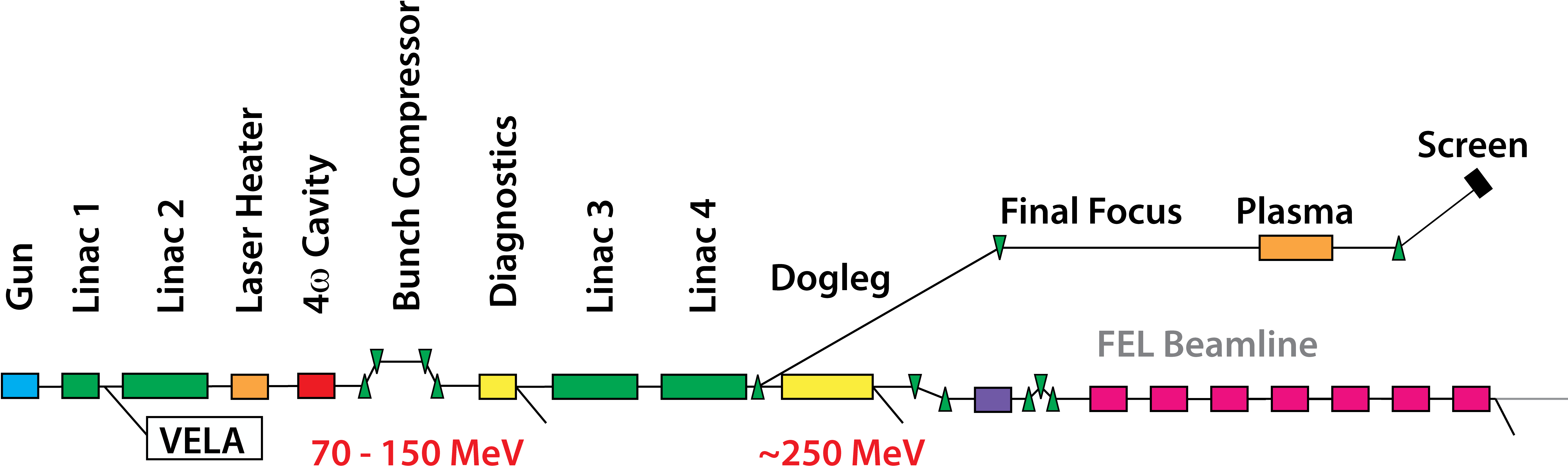}
\caption{Conceptual layout of the CLARA facility and the PARS beam line.}
\label{fig:CLARA_PARS_layout}
\end{figure}

\section{CLARA facility and PARS beam line}
\label{}

The aim of CLARA is to develop a normal conducting test accelerator able to generate longitudinally and transversely bright electron bunches and to use these bunches in the experimental production of stable, synchronized, ultrashort photon pulses of coherent light from a single pass FEL (free electron laser) with techniques directly applicable to the future generation of light source facilities \cite{clarke1}. The CLARA facility houses a photo-injector electron gun, normal conducting accelerating cavities, magnetic bunch compressor, fourth harmonic lineariser, dedicated beam diagnostic sections at low and high energies and FEL beam line, as illustrated in Fig. \ref{fig:CLARA_PARS_layout}. The beam energy is 250$\,$MeV and the maximum bunch charge is about 250$\,$pC. The achievable bunch parameters in different operating modes are listed in Table 1 \cite{clarke2, clarke3}. 

For the electron beam driven PWFA experiment at PARS, a dogleg will guide the full energy CLARA beam to a parallel beam line, offset by $\sim$1.5 m from the CLARA beam axis, contained within the CLARA shielding area. The conceptual layout of the PARS beam line is also shown in Fig. \ref{fig:CLARA_PARS_layout}. It consists of the dogleg beam line, final focus, plasma cell, energy spectrometer (phosphor screen) and final beam dump (not shown). The dogleg beam line consists of arrays of dipoles and quadrupoles to guide and focus the beam from the CLARA beam line to the PARS. The final focus, which is prior to the plasma cell, is designed to focus the electron beam transversely and to match the electron beam with the plasma. A variable 10-50 cm plasma cell (a DC discharge plasma source seems feasible) will be built to test the key issues in the PWFA experiments at various beam and plasma parameter ranges. An energy spectrometer, together with a phosphor screen is employed to characterize the energy of electrons exiting the plasma cell. The final beam dump will absorb the energy of electrons after exiting the plasma cell. Prior to the final focus and plasma cell, a magnetic chicane may be needed to compress the bunch further to an extremely short length. 

\begin{figure}[ht] 
\centering
\includegraphics[width=0.48\textwidth]{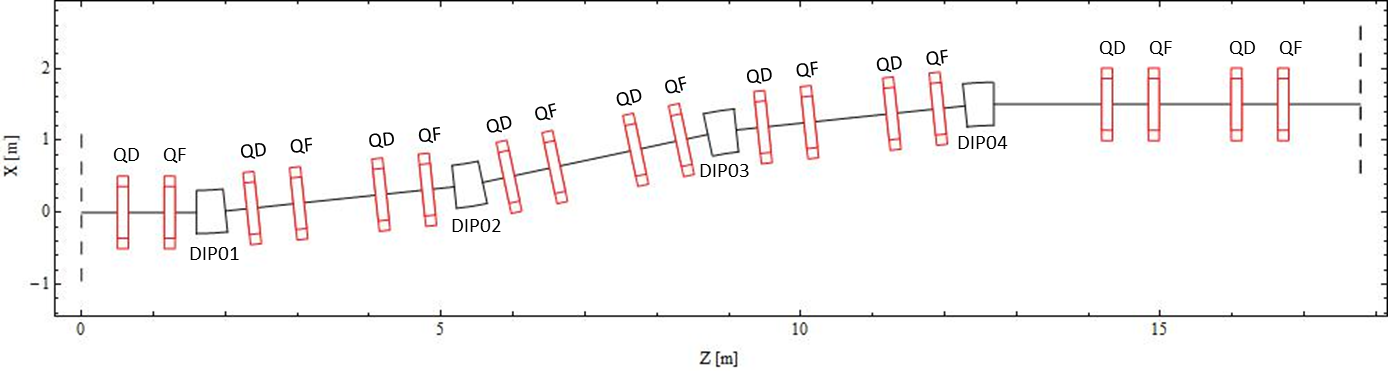}
\caption{The beam line design for PARS by using 4 identical dipoles.}
\label{fig:pars_beamline}
\end{figure}

\begin{figure}[ht] 
\centering
\includegraphics[width=0.35\textwidth]{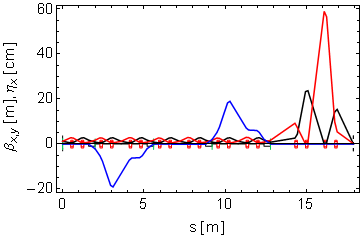}
\caption{Twiss ($\beta_x$ in red and $\beta_y$ in black) and dispersion functions ($\eta_x$ in blue) of the PARS beam line.}
\label{fig:optics_pars}
\end{figure}

\begin{figure}[ht] 
\centering
\includegraphics[width=0.35\textwidth]{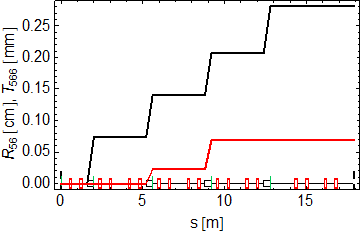}
\caption{The $R_{56}$ (black) and $T_{566}$ (red) of the PARS beam line.}
\label{fig:matrix_ele_pars}
\end{figure}

\begin{figure}[!ht] 
\centering
\includegraphics[width=0.35\textwidth]{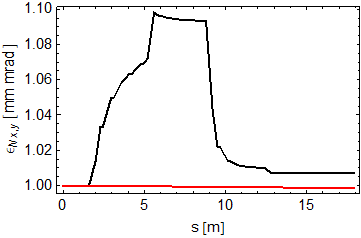}
\caption{Projected normalized beam emittance (black-horizontal, red-vertical) for a Gaussian beam tracked through the PARS beam line (CSR included).}
\label{fig:emitt_pars}
\end{figure}

The proposed dogleg beam line design, using $-I$ transform between the dipoles using two FODO doublets, keeps the transverse beam emittance blow up due to coherent synchrotron radiation (CSR) within acceptable limits \cite{zholents, dimitri}. Figure \ref{fig:pars_beamline} illustrates the beam line design for PARS by using 4 dipoles, with each dipole bending angle of 6$^0$. The optical functions are plotted in Fig. \ref{fig:optics_pars}, from which we can see the transverse dispersion function is fully suppressed at the straight section prior to the plasma cell. Figure \ref{fig:matrix_ele_pars} gives the first order longitudinal dispersion $R_{56}$ and the second order longitudinal dispersion function $T_{566}$ with respect to the PARS beam line axis. Through it, the bunch can be compressed to an ultrashort pulse which is crucial to a high plasma wakefield excitation. Using the typical CLARA beam parameters, e.g. a bunch length of 75$\,\mu$m, a bunch charge of 250$\,$pC and an energy spread of ~1\%, we track a Gaussian beam through the PARS beam line using ELEGANT \cite{borland}. The result is shown in Fig. \ref{fig:emitt_pars}. The beam emittance growth (about 1\%) is negligible before it reaches the plasma cell. It shall be noted that the small emittance change is mainly due to the CSR effects. 

\begin{figure*}[ht]
\centering
\subfigure[]{%
\includegraphics[width=0.4\textwidth]{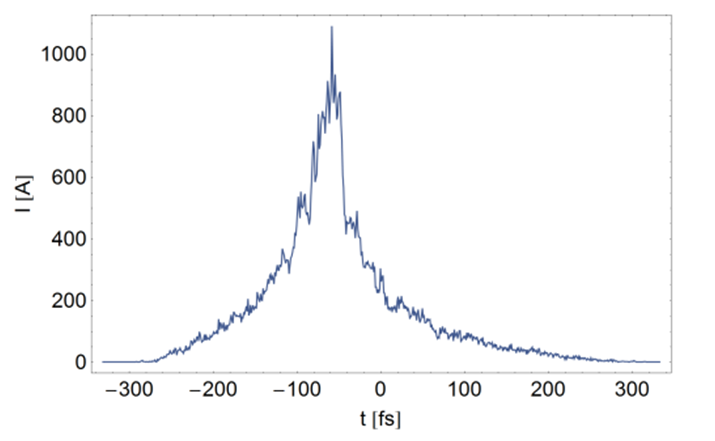}
\label{fig:subfigure1}}
\quad
\subfigure[]{%
\includegraphics[width=0.4\textwidth]{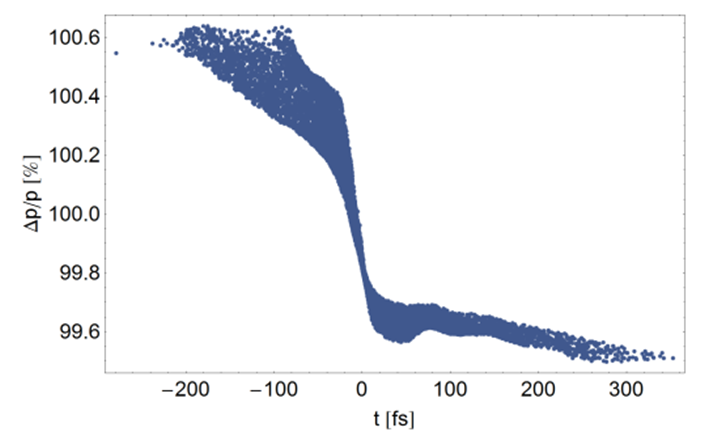}
\label{fig:subfigure2}}
\caption{Longitudinal bunch ((a) current, (b) momentum) distribution at the exit of Linac 2 for a 100$\,$pC electron bunch, including 3D space-charge forces, and tracked with 100$\,$k particles in ASTRA.}
\label{fig:linac2_long}
\end{figure*}


\begin{figure*}[ht]
\centering
\subfigure[]{%
\includegraphics[width=0.4\textwidth]{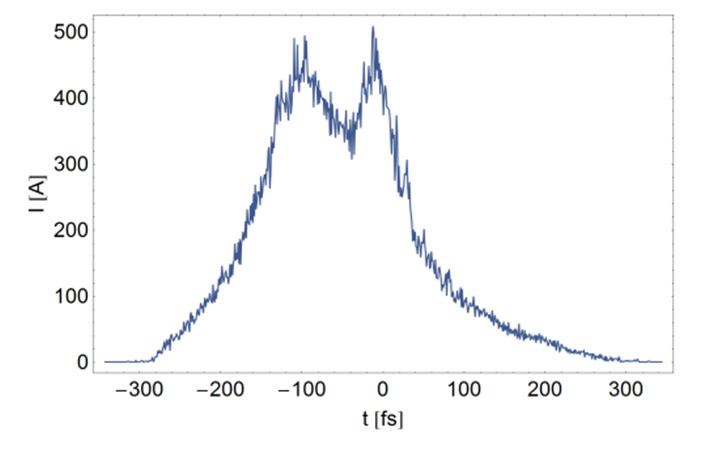}
\label{fig:subfigure1}}
\quad
\subfigure[]{%
\includegraphics[width=0.4\textwidth]{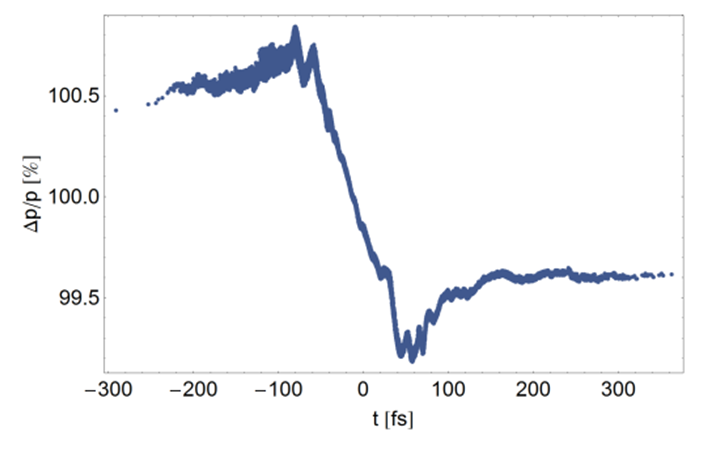}
\label{fig:subfigure2}}
\caption{Longitudinal bunch ((a) current, (b) momentum) distribution at the exit of Linac 4 for a 100$\,$pC electron bunch at 250$\,$MeV and including longitudinal space charge forces using the Elegant code for 100$\,$k particle bunch.}
\label{fig:linac4_long}
\end{figure*}


\begin{figure*}[ht]
\centering
\subfigure[]{%
\includegraphics[width=0.4\textwidth]{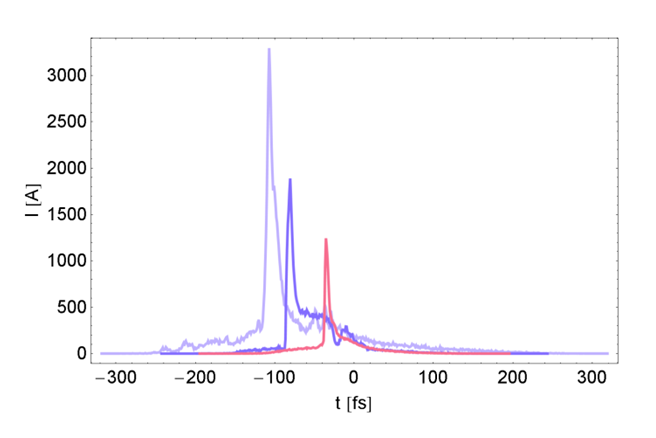}
\label{fig:subfigure1}}
\quad
\subfigure[]{%
\includegraphics[width=0.4\textwidth]{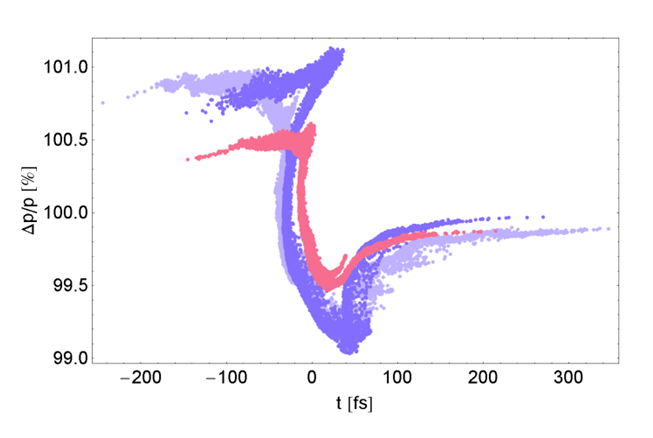}
\label{fig:subfigure2}}
\caption{Final longitudinal bunch properties ((a) current, (b) momentum distribution) at the exit of the PARS beamline for 3 different bunch charges (light blue � 100$\,$pC; dark blue � 50$\,$pC; pink � 20$\,$pC) at 250$\,$MeV and for 100$\,$k particle bunches tracked using Elegant with CSR and longitudinal space-charge forces included.}
\label{fig:final_bunch_PARS_exit}
\end{figure*}


\section{Electron beam tracking at CLARA/PARS}
\label{}

To qualify the tracking simulations performed using a Gaussian bunch profile, full start-to-end simulations were performed from the electron source to the end of the PARS final focus line, using an ASTRA+ELEGANT computational framework. Baseline CLARA design introduces the VELA photo-injector where a Ti-sapphire laser (266$\,$nm) illuminates a copper cathode with 76$\,$fs pulses (1mJ energy per pulse) at a repetition rate of 10$\,$Hz. The cathode is followed by a 2.5 cell, S-band, normal conducting cavity with an accelerating gradient of 100$\,$MV/m \cite{clarke3}. Consequently, the beam is generated at the cathode using the ASTRA code \cite{flottmann}, and the generated beam is tracked through the gun using full space charge tracking up to the first CLARA Linac where it reaches an energy of 6.5$\,$MeV. There are two envisaged operating modes of CLARA. The first mode is a �long�-pulse mode, where the beam is accelerated at some relatively small off-crest phase in the first Linac, and which is magnetically compressed in the down-stream bunch compressor; and the second mode is an �ultra-short�-pulse mode where the beam is velocity-bunched (VB) in the first Linac, and no magnetic compression is performed. The magnetic compression scheme is designed to operate at up to 250$\,$pC. Whereas the velocity bunched mode is limited to bunch charges of 100$\,$pC.

To investigate the limits of achievable bunch lengths for the PARS experiment, initial work has focused on the second option, the velocity-bunched scheme, as this is expected to produce the shortest bunches from CLARA. For this scheme to work the beam must pass through the first Linac around the zero-crossing phase of the RF waveform, that is $\sim$90$^0$ from the crest. At this phase the head of the bunch is decelerated, whilst the tail is accelerated, which, due to the low energies involved, leads to a velocity-time correlation over the bunch. Over a subsequent drift distance, up to the entrance of Linac 2, the head and the tail of the bunch compress due to the velocity correlation and the bunch can reach peak currents of over 2$\,$kA at 100$\,$pC bunch charge. The beam is accelerated in Linac 2 up to 75$\,$MeV, which limits the growth in bunch properties, primarily due to space-charge forces, in the rest of the machine. This section of the machine is also tracked in ASTRA with full space-charge forces included, however the wakefields in Linacs are not included. The output of the ASTRA simulations are then passed to the ELEGANT code, which includes longitudinal space-charge forces as well as a 1-D model of CSR. Representative wake-field models for all RF cavities after Linac 2 are also included in ELEGANT. The beam is tracked through the diagnostics and bunch-compression (switched off in this scenario) sections of the CLARA machine before being further accelerated in Linacs 3 and 4 up to the design energy of 250$\,$MeV. Due to the high-density longitudinal structure of the electron bunch, space-charge forces are still important at the intermediate energies, and this leads to a broadening of the electron bunch longitudinally, which reduces the peak currents at the exit of Linac 4 to approximately 600$\,$A for a 100$\,$pC bunch. Tuning of the exact RF phases in Linacs 1 and 2 can mitigate some of this reduction, but maximum peak currents are always less than 1$\,$kA at 100$\,$pC. The VB mode produces a slight longitudinal chirp in the electron bunch which, due to the very short bunch lengths, is not significantly modified by the off-crest phases in Linacs 2-4. This natural chirp leads to a longitudinal bunch shortening in the PARS beam line, due to the inherent $R_{56}$ as shown in Fig. \ref{fig:matrix_ele_pars}. This $R_{56}$ can be used to increase the peak current at the PARS final focus system to over 3$\,$kA at 100$\,$pC. CSR effects can also be seen in the bunch, but due to the $-I$ phase advance designed into the system, they do not strongly affect the longitudinal bunch properties. The fully longitudinally compressed bunch is then transversely focused in the final-focus system to produce transverse beam dimensions of the order of 20$\,\mu$m. Transverse focusing at higher bunch charges is significantly harder than at lower charges due to the CSR effects that are seen. The transverse beam properties are also affected by space-charge in the long transport from the exit of Linac 2 till the entrance of Linac 3. The current ELEGANT model does not include transverse space-charge forces, but initial estimates using ASTRA do show an increase in the transverse emittance by a factor of 3. The effects of transverse space-charge will be more fully characterised at a later date, and the current resultstherefore represent an idealised case.

The results of these simulations are demonstrated, for a 100$\,$pC CLARA bunch. In Fig. \ref{fig:linac2_long} we show the beam current and momentum distribution at the exit of Linac 2, where the electron beam has been velocity bunched in Linac 1, and then accelerated approximately on-crest in Linac 2. This also corresponds to the beam distribution at the end of the ASTRA tracking run, and thus the input for ELEGANT.

In this simulation we have set Linac 1 at a phase of 91.5$^0$, slightly off from the nominal zero-crossing phase, and Linac 2 at an off-crest phase of + 4$^0$. This acts to limit the velocity bunching and gives peak currents of $\sim$1$\,$kA at the exit of Linac 2, compared to the $\textgreater$2$\,$kA that are possible, for a 100 pC bunch. This reduced peak current at the exit of Linac 2 has been shown to increase the achievable peak currents seen at the end of the PARS beamline by up to 2$\,$kA, by both reducing the effects of space-charge forces in the long transport system up to Linacs 3 and 4, and increasing the longitudinal chirp on the bunch.
    
Figure \ref{fig:linac4_long} shows the same bunch at the exit of Linac 4, with an energy of 250$\,$MeV. Here, the peak current has been reduced to only $\sim$500$\,$A due to the broadening effects of longitudinal space-charge forces. The natural chirp on the bunch is now much more clearly defined, and this will lead to a longitudinal bunch compression in the PARS beamline. There are also 2 clear tails on the bunch distribution at the head and tail of the bunch. These tails arise within from the electron source, and it does not seem possible to mitigate those using solely electro-magnetic means in the beamline. The final longitudinal bunch properties at the exit of the PARS line are shown in Fig. \ref{fig:final_bunch_PARS_exit}. Tracking results for different bunch charges of 20$\,$pC and 50$\,$pC are also presented in comparison with the results for 100$\,$pC.

\begin{figure}[ht] 
\centering
\includegraphics[width=0.38\textwidth]{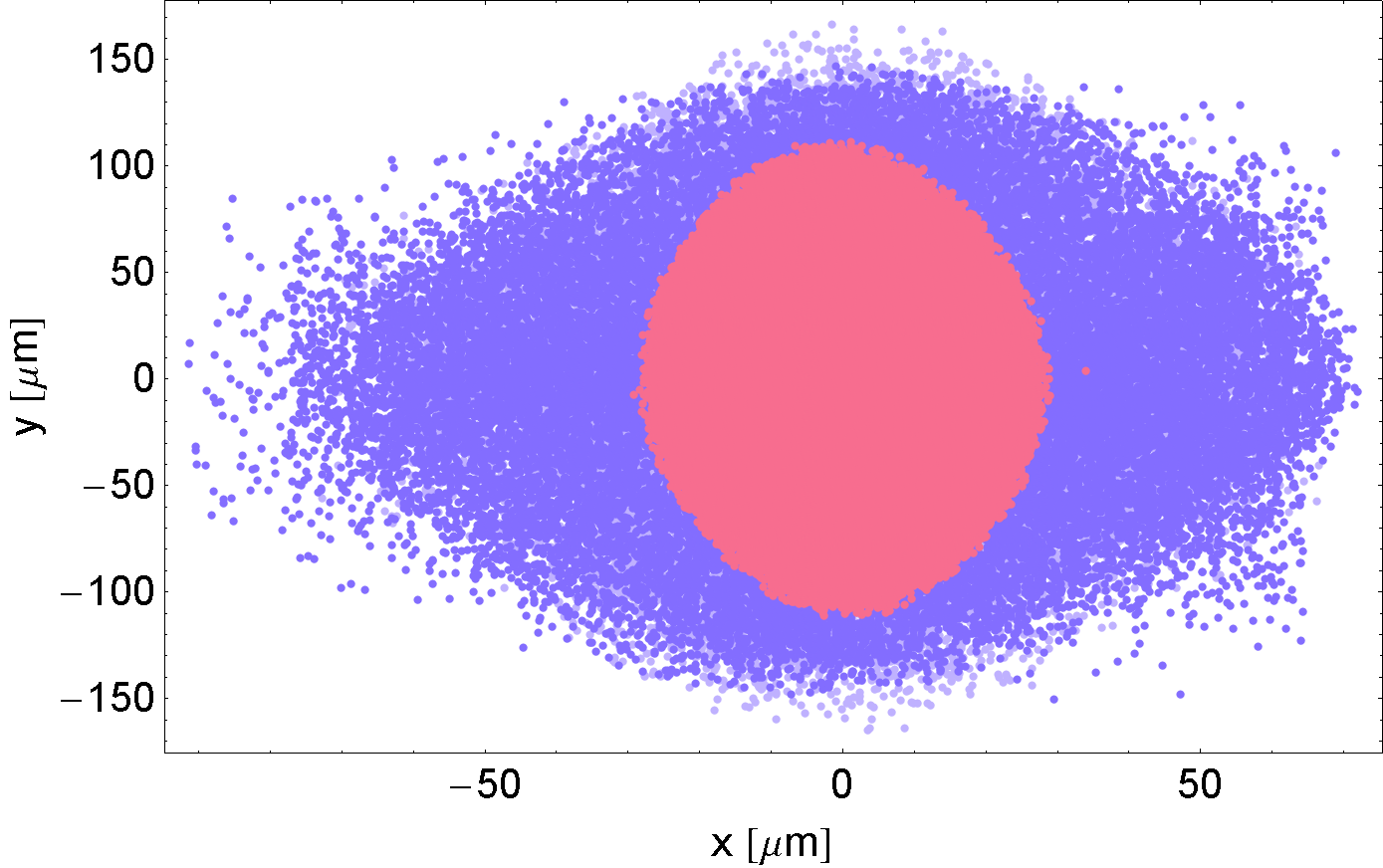}
\caption{Transverse beam sizes at the exit of the PARS beamline for three different bunch charge values (light blue - 100$\,$pC; dark blue - 50$\,$pC; pink - 20$\,$pC) at 250$\,$MeV and for 100$\,$k particle bunches tracked using Elegant with CSR and longitudinal space-charge forces included. 100$\,$pC and 50$\,$pC distributions are almost identical.}
\label{fig:pars_spot_end}
\end{figure}

Figure \ref{fig:final_bunch_PARS_exit} clearly shows the high expected peak currents of $>\,$3$\,$kA at 100 pC, but also peak currents of $\sim$2$\,$kA at 50$\,$pC and $>\,$1$\,$kA at 20 pC. The non-linear reduction in peak current with reducing charge is mostly due to mitigation of the space-charge broadening in the CLARA beamline at reduced bunch charges. There is also some reduction in the CSR effects in the PARS beamline itself, although this is less important for the final longitudinal bunch length. Due to the bunch tails that arise from the electron source, there is a broad background of charge in the longitudinal bunch distribution. This limits the total charge in the current spike to, at worst, 50$\%$ of the main bunch charge.

In all three cases the bunch length for the current spike is shorter than 30$\,$fs RMS. At the exit of the PARS beamline, the bunch has to be focused transversely as well as longitudinally. Figure \ref{fig:pars_spot_end} shows the transverse bunch properties for the three different charge values shown above.

It can be seen that for a low bunch charge the transverse beam size falls to about 20$\,\mu$m RMS size. However, for high charge regimes, 50$\,$pC and 100$\,$pC, there is some broadening of the bunch transverse size due to the increased energy spread of the beam, arising due to CSR effects in the PARS beamline. Note, however, that the increase in transverse emittance due to space-charge forces has not yet been fully characterised, and these numbers are expected to increase.

\begin{table*}[t] 
\caption{Three operation regimes for the PWFA experiment at the CLARA/PARS facility.} 
\centering 
{\small
\begin{tabular}{l c c c c} 
\hline\hline 
Operating modes & Long Pulse & Short Pulse & Ultra-Short Pulse \\ [0.5ex] 
\hline 
Beam energy (GeV) & 250 & 250 & 250  \\ 

Charge/Bunch Q (pC) & 250 & 250 & 20-100  \\

Electron/Bunch $N_b$ ($\times10^9$) & 1.56 & 1.56 & 0.125-0625 \\

Bunch length rms (fs) & 250-800 (flat top) & 100-250 & $\leq$ 30 \\

Bunch length ($\mu$m) & 75-240 & 30-75 & 9 \\

Bunch radius  ($\mu$m) & 20-100 & 20-100 & 20-100 \\

Normalised emittance (mm$\,$mrad)& $\leq$ 1 & $\leq$ 1 & $\leq$ 1 \\

Energy spread (\%) & 1 & 1 & 1 \\  [1ex] 
\hline 
\end{tabular} }
\label{table:nonlin} 
\label{tab:beam_specs}
\end{table*}

\section{PARS beam parameters}
\label{}

According to the linear theory of PWFA \cite{lee}, the maximum plasma wakefield amplitude is proportional to the bunch charge (number of electrons per bunch) and inversely proportional to the square of the bunch length. Therefore a short drive beam is in principle preferable for a high amplitude wakefield excitation. A preliminary simulation study has shown that an accelerating gradient of a few GV/m can be obtained if a short (therefore high peak current) electron bunch is used as drive beam at the PARS.

Based on the initial tracking results for VB mode mentioned above, as well as the details found in the CLARA CDR \cite{clarke3}, we have defined the parameter settings for the PWFA experiment at the PARS beam line \cite{clarke3}. Table 1 summarizes the different working regimes of the beam parameters; each parameter setting can be used to study the beam-plasma interactions and test the scaling laws of PWFA. In principle three operating regimes, i.e, long pulse, short pulse and ultra-short pulse can be achieved at CLARA facility for the PWFA experimental research \cite{xia1}.

\section{Particle-in-cell Simulations}
\label{}
\subsection{Wakefield structures}
\label{}

\begin{figure}[!ht] 
\centering
\includegraphics[width=0.35\textwidth]{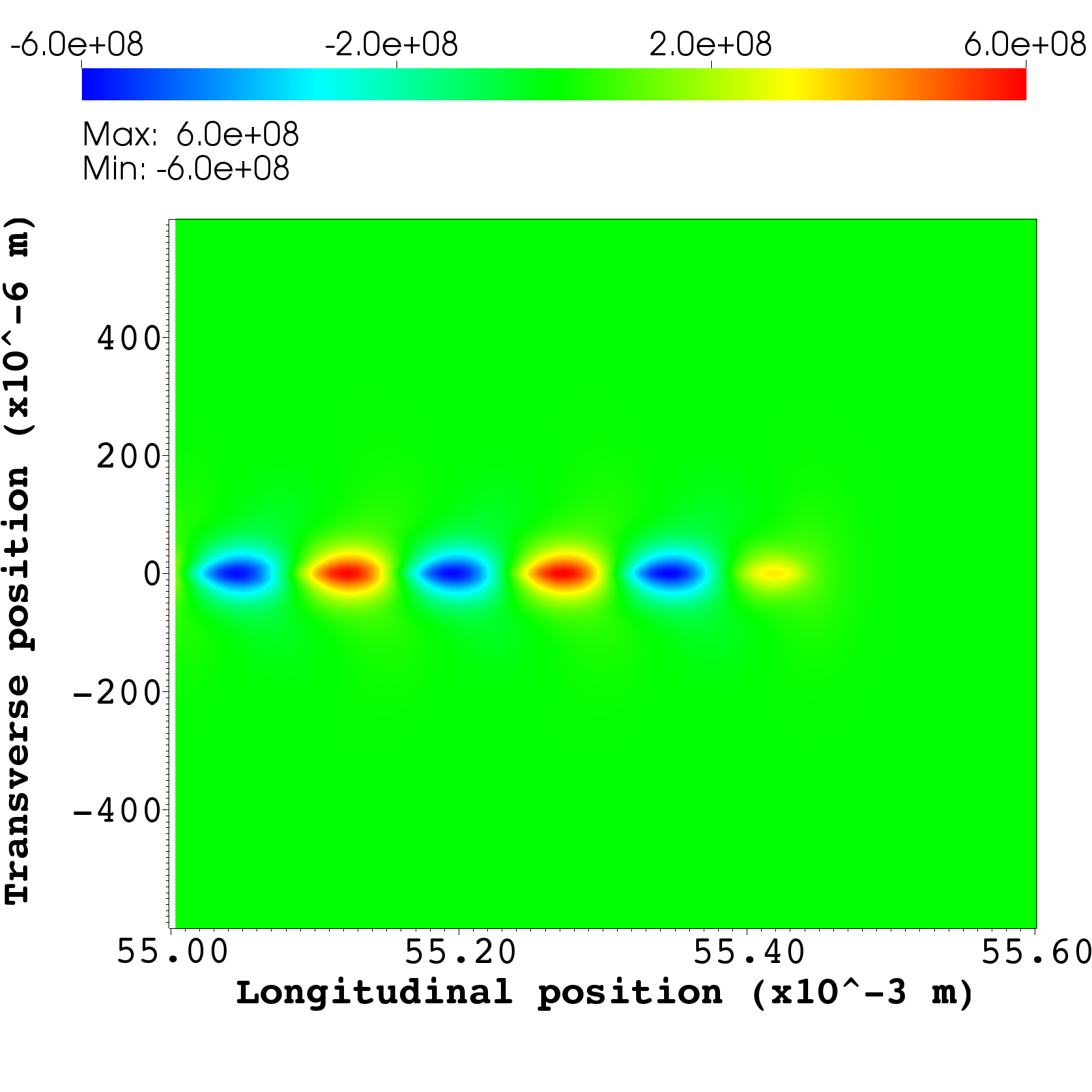}
\caption{The accelerating and decelerating wakefield structures (blue and red regions) driven by a typical CLARA beam.}
\label{fig:wakefield_structures}
\end{figure}

\begin{figure}[!ht] 
\centering
\includegraphics[width=0.33\textwidth]{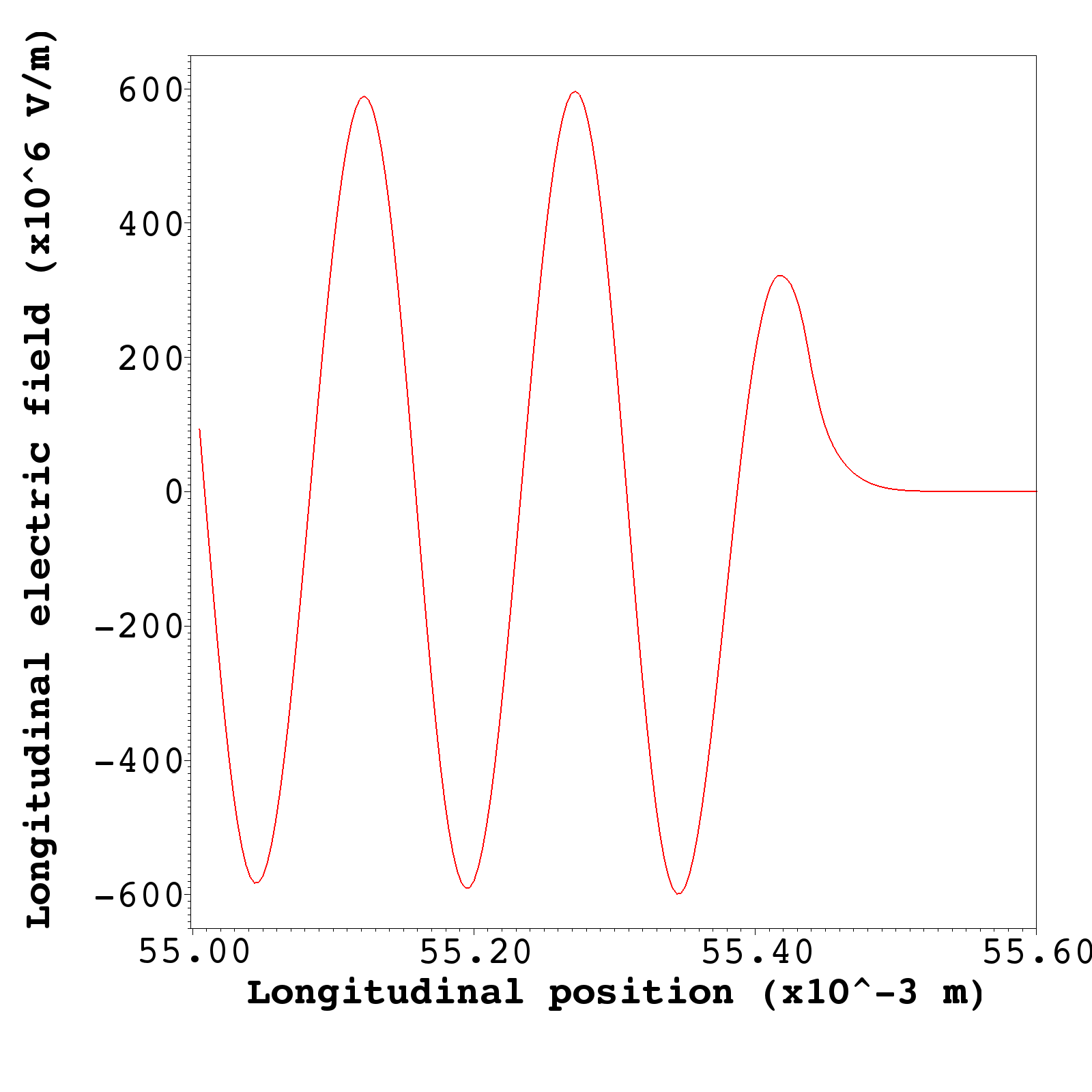}
\caption{Longitudinal accelerating and decelerating wakefields driven by CLARA beam in a plasma. }
\label{fig:efield}
\end{figure}

To gain a full understanding of the wakefield excitation by using the relativistic CLARA beam as drive beam, one has to rely on detailed simulations. VORPAL is a fully explicit PIC code developed by Tech-X Corporation in Boulder, USA \cite{link}. We use it to simulate the interactions between electron beam and plasma. Figure \ref{fig:wakefield_structures} gives a 2D simulation of typical accelerating and decelerating wakefield structure in a plasma. The parameters used in this simulation are as follows: the beam energy is 250$\,$MeV, bunch length is 30$\,\mu$m, transverse beam size is 100$\,\mu$m, bunch charge is 250$\,$pC, and plasma density is $5\times10^{16}\,cm^{-3}$. The maximum longitudinal accelerating and decelerating wakefield $E_x$ is about 600$\,$MV/m, as illustrated in Fig. \ref{fig:efield}, an order of magnitude higher than what can be achieved with conventional RF structures.

\subsection{Effect of plasma density}
\label{}

\begin{figure}[ht] 
\includegraphics[width=0.485\textwidth]{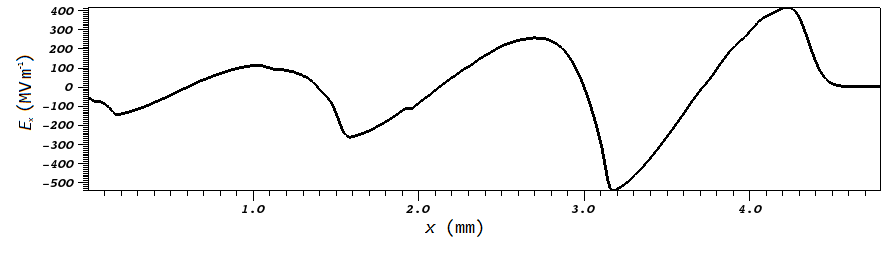}
\caption{Longitudinal wakefields for a plasma with density an order of magnitude less denser than the drive beam density.}
\label{fig:long_wfield1}
\end{figure}

\begin{figure}[ht] 
\includegraphics[width=0.485\textwidth]{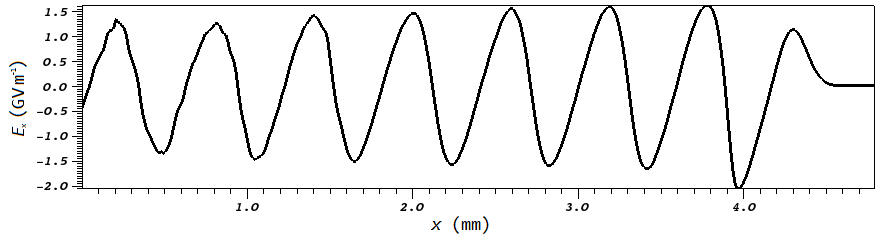}
\caption{Longitudinal wakefields for a plasma with density equal to the drive beam density.}
\label{fig:long_wfield2}
\end{figure}

\begin{figure}[ht] 
\includegraphics[width=0.485\textwidth]{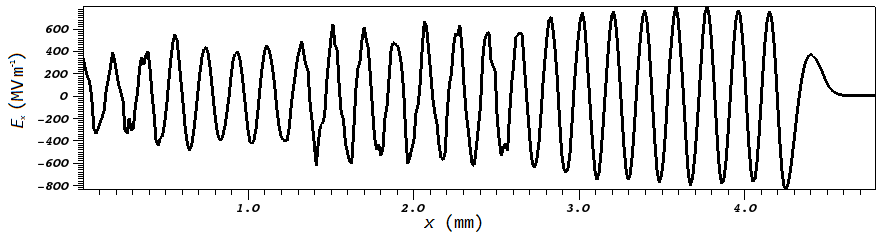}
\caption{Longitudinal wakefields for a plasma with density an order of magnitude denser than the drive beam.}
\label{fig:long_wfield3}
\end{figure}

The effect of plasma densities on the plasma wakefields, while keeping other parameters constant was investigated by using simulations at densities one order of magnitude lower than, higher than, and equal to the electron bunch density, with beam parameters expected at the proposed CLARA beam. Fig. \ref{fig:long_wfield1}, \ref{fig:long_wfield2}, and \ref{fig:long_wfield3} show the longitudinal wakefields $E_x$ for these different situations, in a 1-D plot against $x$. Here, $x$ denotes the longitudinal beam propagation direction. The input parameters are from the short pulse mode as listed in Table 1. The beam radius used is 20$\,\mu$m and the bunch length is 75$\,\mu$m, which corresponds to a beam density of $3.3\times10^{15}\,cm^{-3}$. 

The simulation results show how the magnitude of the plasma wakefields depends on the plasma densities, and that the accelerating field is modulated at the plasma wavelength $\lambda_p$. The highest accelerating gradient is achieved for a plasma density matching the electron beam density. It is also notable that the wakefield amplitude decreases most rapidly for the low density plasma, and least rapidly for the plasma equal in density to the electron beam. The maximum accelerating field is about 2$\,$GV/m when the plasma density is the same as the drive beam density.

\subsection{Strongly focused beam as driver}
\label{}

\begin{figure}[ht] 
\centering
\includegraphics[width=0.36\textwidth]{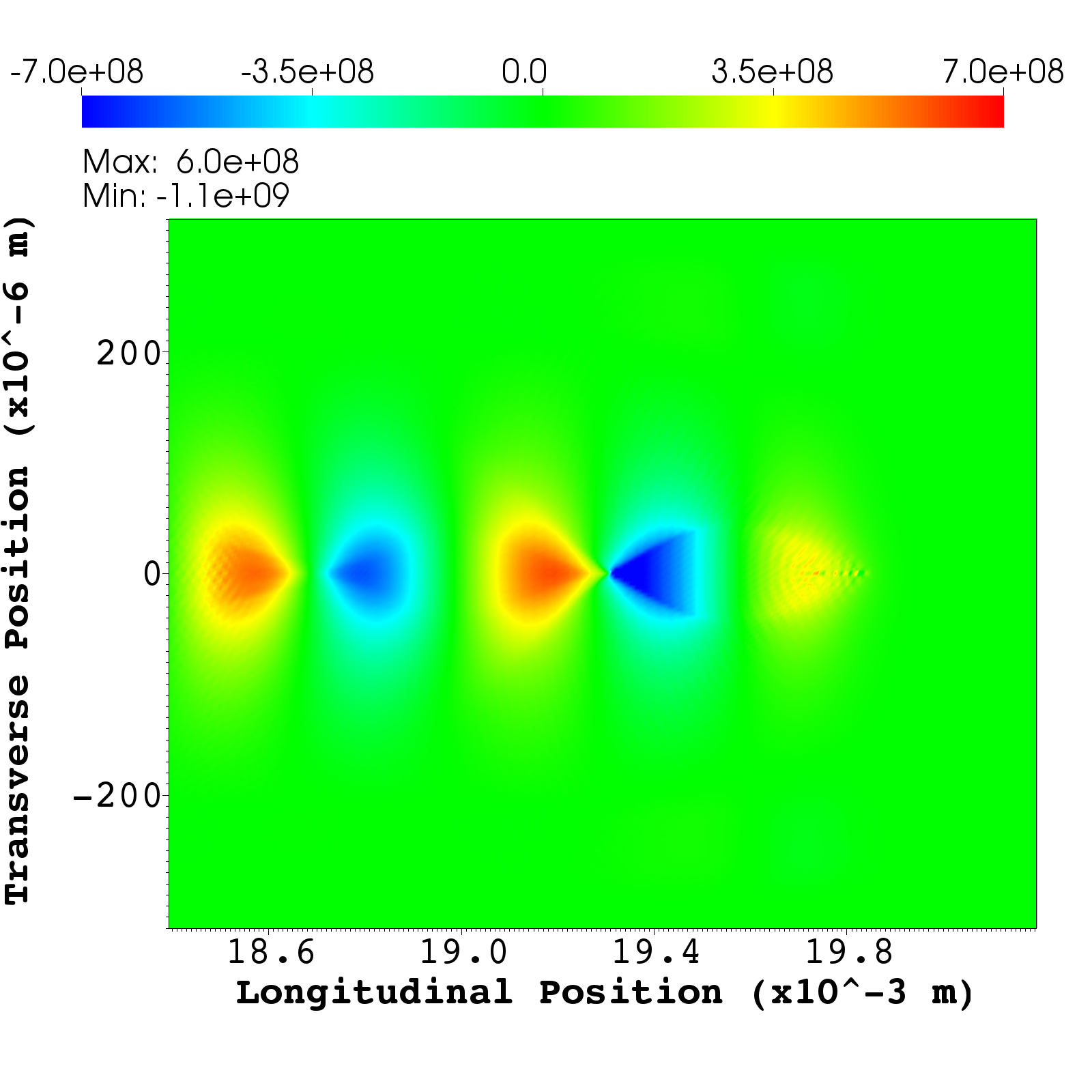}
\caption{The accelerating and decelerating wakefield structures driven by a strongly focused CLARA beam.}
\label{fig:wakefield_structures2}
\end{figure}

\begin{figure}[ht] 
\centering
\includegraphics[width=0.34\textwidth]{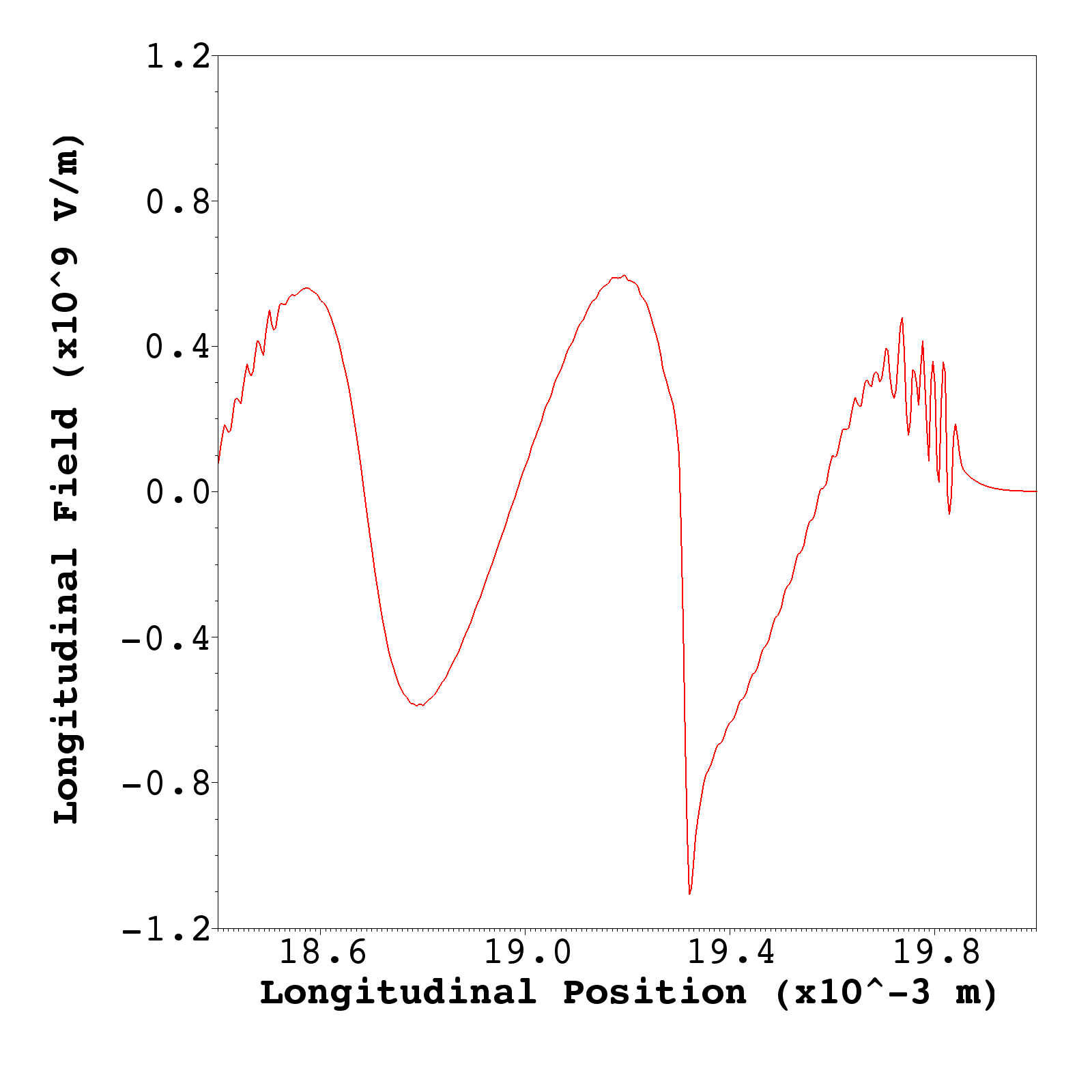}
\caption{The longitudinal wakefields driven by CLARA strongly focused beam.}
\label{fig:wfield}
\end{figure}

If the electron beam is subject to strong transverse focusing, i.e., by final focus prior to the plasma cell, one expects a high-density drive bunch can be achieved. The bunch density for a Gaussian distributed beam is given by

\begin{equation}
n_b = N_b / \Big[  (2\pi)^{3/2} \sigma^2_r \sigma_z  \Big]
\label{eqn:bunch_density}
\end{equation}
where $N_b$ is the number of electrons per bunch, $\sigma_r$ and $\sigma_z$ are the transverse RMS beam sizes and the bunch length, respectively. It can be seen that the bunch density is proportional to the number of electrons per bunch and inversely proportional to the bunch length and the square of the transverse RMS beam size. Figures \ref{fig:wakefield_structures2}-\ref{fig:wfield} give the simulation results based on a strongly focused CLARA beam driven wakefields. In this simulation, the electron beam energy is 250$\,$MeV, the transverse beam size is 20$\,\mu$m, the bunch length is 75$\,\mu$m, the bunch charge is 250$\,$pC and the normalized emittance is 1$\,$mm.mrad. The plasma density is $3\times10^{15}\,cm^{-3}$. Figure \ref{fig:wakefield_structures2} shows the delicate accelerating and decelerating structures driven by CLARA beam. The longitudinal wakefield amplitude is shown in Fig. \ref{fig:wfield}. One can see that in this case, the maximum accelerating field is approaching $\sim$1.15$\,$GV/m.  

Based on linear theory of PWFA, the maximum wakefield amplitude is given by \cite{lee}

\begin{equation}
E_{acc}[MeV/m] = 244 \frac{N_b}{2\times10^{10}} \Bigg( \frac{600\mu m}{ \sigma_z (\mu m)} \Bigg)^2.
\label{eqn:E_acc}
\end{equation}

It gives a maximum wakefield amplitude of about 1.2 GV/m by using the strongly focused beam parameters, which agrees well with the simulation results as shown in Fig. \ref{fig:wfield}.

\begin{figure}[ht] 
\centering
\includegraphics[width=0.4\textwidth]{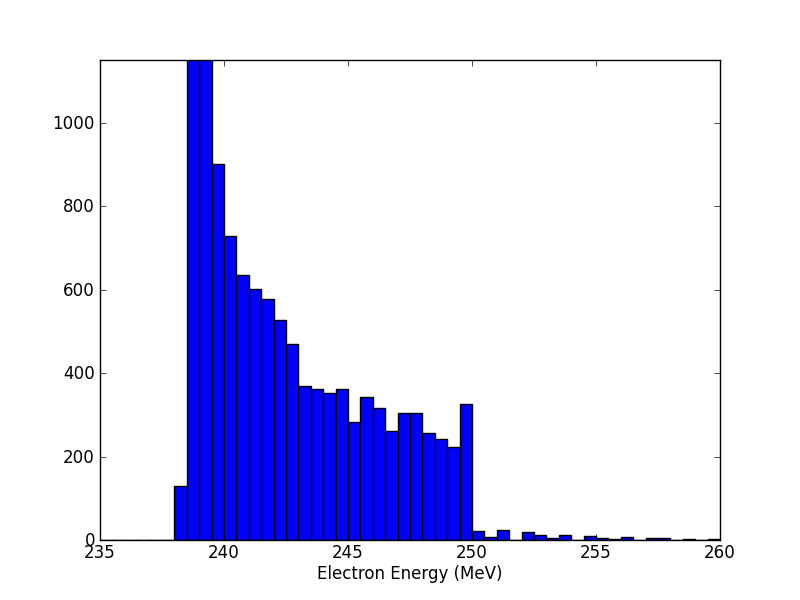}
\caption{The beam energy distribution after propagating through a 3 cm plasma cell for a strongly focused CLARA drive beam.}
\label{fig:energy_hist}
\end{figure}

Meanwhile, one can also examine the particle acceleration effect from the CLARA beam driven plasma wakefield. Figure \ref{fig:energy_hist} shows the 3D simulation result on beam energy spectrum after propagating through a 3$\,$cm long plasma cell for a strongly focused CLARA drive beam, with the same beam parameters as used in Figure \ref{fig:wakefield_structures2}.  It can be seen that most of the particles in the bunch lose their energy into the plasma, while a small amount of particles in the bunch gain energy from the plasma.

\section{Other research topics at PARS}
\label{}

There are many interesting research topics which can be explored using the CLARA beam driven PWFA. For example, a two-bunch acceleration experiment, i.e., one bunch for driving the wakefield and the other bunch for sampling the wakefield, can be studied at PARS facility. Based on the latest simulations, one could hope to double the energy of the CLARA beam with a 20$\,$cm preformed plasma. In this case, the FEL seeded from the wakefield accelerated beam will enable short wavelength photon production for scientific research and industrial users.

In addition, the plasma can act as an undulator to produce a high brightness photon beam at energies between keV to MeV through betatron radiation \cite{wang2}. The mechanism of the radiation production can be extensively investigated at PARS facility.

Furthermore it is possible to study the self-modulation instability (SMI) of a long electron beam in a high-density plasma. The AWAKE collaboration at CERN will investigate self-modulated SPS proton driven wakefield acceleration \cite{muggli2, xia2}. The CLARA beam has the same gamma factor as the SPS beam, and it is much easier to manipulate than the SPS beam due to its lower energy. Therefore the experimental results from CLARA/PARS will be able to give some inputs to the CERN AWAKE experiment.

Except for the aforementioned plasma wakefield acceleration experiments, a dedicated beam line at PARS will allow us to study dielectric wakefield acceleration which is currently under study at the Cockcroft Institute \cite{aimidula} and Daresbury laboratory \cite{saveliev}. Meanwhile the beam from PARS can be used to test other advanced beam dynamics issues, e.g. coherent synchrotron radiation and its countermeasures, microbunching instability, emittance exchange and some novel FEL schemes. The versatile PARS beam operation modes make it an ideal test bench for beam instrumentation R$\&$D, e.g. the coherent Smith-Purcell radiation, coherent edge radiation and electron-optical sampling (EOS) etc. 

\section{Conclusions}
\label{}

The proposed PARS at CLARA facility will address the key issues in electron beam driven plasma wakefield acceleration. The PARS beam line has been designed and it has demonstrated that three sets of beams, e.g. long pulse, short pulse and ultrashort pulse, are feasible in the beam transport from the CLARA to the PARS. The particle-in-cell simulations show that the high accelerating gradients from a few hundreds of MV/m to several GV/m can be achieved from the plasma wakefields. In addition, this specially designed PARS beam line can be used to study other advanced beam dynamics issues and enable many beam applications.

\section*{Acknowledgements}
\label{}

The simulations in this paper were conducted at the STFC Hartree Supercomputer Centre. We acknowledge their great help for simulations. This work is supported by the STFC and the Cockcroft Institute Core Grant.







\bibliographystyle{unsrt}

\end{document}